\documentstyle[12pt] {article}
\begin {document}
\global\parskip 6pt

\newpage
\parindent=0pt
\begin {center}
{\large\bf Microscopic interface phonon modes in structures of
GaAs quantum dots embedded in AlAs shells}\\
\vspace{0.6cm}
Shang-Fen Ren and G. Qin$^{\dagger}$ \\
Department of Physics, Illinois State University,
Normal, IL 61790-4560\\
\vspace{0.8cm}
(Received $\hspace{8.2cm}$ )\\
\vspace*{8.5cm}
\end {center}
pacs {63.22.+m, 81.05.Ys, 81.05.Ea, 81.40.Tv}
\vspace{1.0cm}

\begin {minipage}{15.0cm}
\hspace*{0.8cm}

By means of a microscopic valence force field model, a series of novel
microscopic interface phonon modes are identified in shell quantum dots
(SQDs) composed of a GaAs quantum dot of nanoscale embedded in an
AlAs shell of a few atomic layers in thickness. In SQDs with such thin
shells, the basic principle of the continuum dielectric model and
the macroscopic dielectric function are not valid any more. The
frequencies of these microscopic interface modes lie inside the gap
between the bulk GaAs band and the bulk AlAs band, contrary to the
macroscopic interface phonon modes. The average vibrational energies
and amplitudes of each atomic shell show peaks at the interface
between GaAs and AlAs. These peaks decay fast as their
penetrating depths from the interface increase.\\


\end{minipage}

\newpage
\hspace*{0.8cm}
\section*{1. Introduction}Interesting results have been reported on the surface
and interface modes of phonons and polaritons in layered or spherical
structures by means of macroscopic continuum model. By using the
Rosenzweig model, G. Armand et al. \cite{Armand} have found that
there are surface phonon modes in the gap of the bulk phonon modes
and below the lowest bulk band. Maradudin et al.\cite{Maradudin}
have predicted that there are surface phonon waves propagating in
the gap between the bulk phonon bands and below the lowest bulk
phonon band in structures that a semi-infinite GaAs/AlAs
superlattice is in contact with a thin film of GaAs or AlAs. The
amplitudes of some surface modes show strong decaying or variation as
their penetrating depths from the surface increase.
At about the same time, Quinn et al. predicted that there are
surface plasmon modes without suffering Landau damping surviving
below, above or between the bands of bulk plasmon frequencies in
both type-I and type-II semi-infinite semiconductor
superlattices\cite{Giuliani, Quinn}. Interface and surface modes
in shell quantum dots (SQDs) composed of a spherical core of one
material (core material) embedded in a matrix of another
materiel (shell material) have been studied by many authors
\cite{Cardona, Ruppin, Ruppin2, Reinecke}. However,
most of the theoretical treatments on shell QDs are
performed in the framework of mechanical continuum
model \cite{Maradudin}, continuum dielectric model\cite{Quinn, Ruppin2},
or continuum model coupling both the mechanical vibrational amplitudes
and the electrostatic potential \cite{Cardona}. Their regions of
validity is limited to modes whose effective wavelength is large
compared to the interatomic spacing. In the case of surface plasmons
the continuum dielectric approach is limited to dots whose size is
large enough that the plasma frequency greatly exceeds the
differences between the single-particle energies \cite{Reinecke}.
After surveying systematically the validity of the dielectric
approximation in describing electron-energy-loss spectra of
surface and interface phonons in thin films of ionic crystals,
Lambin et al. \cite{Lambin} demonstrated that the dielectric
approximation reproduces the essential features of the phonon
response when the layer thickness exceeds 20-30 \AA. 
approximation can no longer be applied to thinner films as
the concept of bulk dielectric function, the only input required in
this approach, breaks down. Even for films in above
thickness range, small contributions of microscopic surface
phonons survive and they may not be neglected. The eigenvectors of the microscopic surface phonons are found to be large at the first surface layer and rapidly decreasing as the distance from the surface increases.   
Therefore to describe phonon modes in semiconductor SQDs with small dots size
and thin shells, an anisotropic microscopic model is
especially needed. \\

\section*{2. The theoretical formalism}

We have developed a microscopic valence force field model (VFFM)
in recent years to investigate phonon modes in QDs \cite{Ren,
Ren2,Jap}.  In this model, the change of the total energy
due to the lattice vibration is considered as two parts,
\begin{math}
\Delta E = \Delta E_{s} + \Delta E_{c}
\end{math},
where the energy changing due to short-range interactions
describes the covalent bonding, and the long range part
approximates the Coulomb interactions
\cite{Kunc,Ren3,Ren4}. For the short range part,
we employed a VFFM as\cite{Harrison},
\begin{math}
\Delta E_{s}=\sum\limits_{i}\frac{1}{2}C_{0} (\frac {\Delta
d_{i}}{d_{i}})^2 +\sum\limits_{j}\frac{1}{2} C_{1} (\Delta \theta_{j})^2
\end{math},
where $C_{0}$ and $C_{1}$ are two parameters to describe the
energy change due to the bond length and the bond angle
respectively, and the summation runs over all the bond
lengths and bond angles. Because each of these two parameters
has a simple and clear physical meaning, this model allows us
to treat the interaction between atoms at the surface and
interface appropriately.

In our model, the projection operators of the irreducible
representations of the group theory is employed
to reduce the computational intensity \cite{SYR,SYR2,SYR3}.
For example, the dynamic matrix for a 8.0 nm GaAs/AlAs SQD is in the
order of 35,565. This can be reduced to five matrices in five different
representations of A$_{1}$, A$_{2}$, E, T$_{1}$, and T$_{2}$, with the
sizes of 1592, 1368, 2960, 4335, and 4560 respectively. This
approach further allows us to investigate phonon modes with different
symmetries in QDs in detail. \\

By employing this model, we have investigated phonon modes in GaAs/AlAs
SQDs composed of a spherical GaAs core of radius d$_{S}$ embedded in
a AlAs shell with external radius d$_{L}$ \cite{Jap}. Our
theoretical formalism has considered every details in the SQDs.
it is suitable to deal with SQDs that the the core material and
the shell material have totally different parameters. But for
calculations of GaAs/AlAs SQDs, the parameters $C_{0}$, $C_{1}$, and
$e^{*}$ for both GaAs and AlAs are taken to be the same for simplicity,
and only the mass difference is considered. The $C_{0}$, $C_{1}$, and
$e^{*}$ are taken as 38.80, 0.858, and 0.6581, and the masses of Ga, As
and Al are taken as 69.723, 74.922 and 26.982 respectively in atom unit.

When considering the interaction between atoms, special attention
is paid to atoms near the surfaces and interface of the shell
QDs. More specifically, for the short range interaction, when
an atom is located near the surface, interaction from its nearest
neighboring atom is considered only if that specific nearest atom
is within the QD, and interaction from its second
neighboring atom is considered only if that specific second neighbor
atom is in the QD as well as the nearest neighboring atom that
makes the link between them.

\section*{3. Results and discussion} By employing this model, the frequencies and
vibrational strengths of phonon modes of GaAs/AlAs SQDs are
calculated as functions of the size of internal dot and the thickness
of the external shell of SQDs in  each of five representations of A$_{1}$,
A$_{2}$, E, T$_{1}$ and T$_{2}$ \cite{Jap}. The results of our
model shows that the entire optical frequency range of SQDs is
divided into two nonoverlaping bands, which are originated from
bulk AlAs band and bulk GaAs band respectively, so named as AlAs-like
band and GaAs-like band respectively. The lowest and highest
frequency of the bulk AlAs (GaAs) optical band is 318.48 $cm^{-1}$ and
396.00 $cm^{-1}$ (268.80 cm$^{-1}$ and 292.13 cm$^{-1}$) respectively
evaluated from the VFFM. We have also noticed that there are many
phonon modes inside the gap between bulk GaAs band and the bulk AlAs
band in SQDs, especially in SQDs covered by thin shells to which the
concept of bulk dielectric function is no longer available.
In this letter, we concentrate our attention on the study of phonon
modes in SQDs with shells of a few atomic layers by using VFFM.

Fig.1 is a plot of density of states of phonon modes for
10 selected 10 symmetries and scales of SQDs with thin shells. It is
plotted by taking half width of the Gaussian broadening
$\sigma=0.4 cm^{-1}$ and the frequency step $\Delta \omega=0.2 cm^{-1}$.
Two thin vertical line is used to show the upper edge of
the bulk GaAs band and the lower edge of the bulk AlAs band.
According to the dielectric model, all the interface modes
corresponding to different angular quantum number ${\it l}$ fall
within the bulk AlAs band or bulk GaAs band respectively,
no matter the shell thickness of the GaAs/AlAs SQD is infinite
or finite (for the latter case, we suppose that the SQD is
enclosed by vacuum). Therefore,  all the modes
with frequencies inside the gap between GaAs and Alas bulk bands
(with 292.13 cm$^{-1} \leq  \omega \leq 318.48$ cm$^{-1}$) should be
categorized into microscopic interface modes, since they are
novel modes that could be revealed only by microscopic models.
Indeed these modes have obvious characteristics of the
interface modes as we will show below. To distinguish with these
microscopic interface modes, we call other modes with frequencies
within the bulk AlAs band or the bulk GaAs band internal modes.

It is seen from Fig. 1 that for QDs with d$_{S}$=21.70
\AA and d$_{L}$ =26.95 \AA (or a 21.70 \AA/26.95 \AA SQD),
there are two T$_{2}$ modes with frequencies of 299.13 cm$^{-1}$
and 304.64 cm$^{-1}$ and a single A$_{1}$ mode with frequency of
305.71 cm$^{-1}$ in the gap between the GaAs-like phonon band
and the AlAs-like phonon band. Similarly in that frequency range,
for a 31.34 \AA/34.26 \AA SQD, there are two modes of
A$_{1}$ and T$_{2}$ symmetries, no mode of A$_{2}$ symmetry,
and one single mode of T${1}$ and E symmetries, respectively.
These are microscopic interface modes that are caused by the
microscopic bonding condition at the interface that completely
fail to survive in macroscopic continuum dielectric model. In
this letter main attention is paid to exam the characteristics
of these microscopic interface modes. \\

\hspace*{0.8cm}
We first exam the average vibrational amplitudes (AVA)
of these atomic shells for the microscopic interface phonon
modes. The AVA of the ${\it l}$-th atomic shell is defined as
\begin{math}
A_l^i=\frac{1}{n}\sum\limits_{k={\it l}}^n |a_{lk}^i|
\end{math},
where $a_{lk}^i$ is the vibrational amplitude of the $k$-th atom in
the $l$-th shell in the $i$-th phonon mode, and $n$ is the total
number of atoms in the ${\it l}-$th shell. In our spherical QD,
th center of the QD is chosen on an atom, and the shells consist
of atoms with the same distance to the center of the QD. Two sets
of plots are shown in Fig. 2 to investigate the behavior the AVA
of microscopic interface modes. These plots show the AVA of atoms
in the $l$-th shell as a function of the shell's diameter.
The mode order number and the frequency of the mode are indicated
in each mode. The upper four subfigures in Fig. 2 are for four
modes of T$_2$ symmetry in a 21.70 \AA/26.95 \AA SQD with the
frequency range from 286.92 cm$^{-1}$ to 331.92 cm$^{-1}$,
while the lower four are for the four modes of A$_1$ symmetry in
a 31.34 \AA/34.26 \AA SQDs with the frequency range from 286.43
cm$^{-1}$ to 330.44 cm$^{-1}$. It is seen that the peaks
of AVA of the 164-th and 165-th modes of T$_{2}$ symmetry with
frequencies of 299.13 and 304.64 cm$^{-1}$ appear exactly at
the vertical thin line which indicate GaAs/AlAs interface
of diameter 21.70 \AA. These are typical microscopic interface
modes. The 163-th mode and the 166-th mode of
T$_{2}$ symmetry have frequencies inside the bulk GaAs and bulk
AlAs bands, so their AVA show the basic characteristics of GaAs
bulk modes and AlAs bulk modes \cite{Ren2}.  Similarly, for
modes of A$_{1}$ symmetry in a 31.34 \AA/34.26 \AA SQD, the AVA
of the 131-th and 132-th modes have peaks at or near the interface
shell indicated by a vertical line. Here we want to add another
important effect: due to the quantum confinement effect discussed in
detail in Ref.s \cite{Ren,Ren2,Jap}, the highest frequency of the
GaAs-like modes decreases and the lowest frequency of the AlAs-like
modes increases as the core size and the shell thickness decrease,
which makes the gap between the top of the GaAs-like band and the
AlAs-like band much wider than the gap between two bulk materials.
Therefore, some modes appear outside the gap but near the gap
may still be microscopic interface modes. The AVA of the 130-th
mode (with frequency of 286.43 $cm^{-1}$) of A$_1$ symmetry for the
31.34 \AA/34.26 \AA SQD is certainly this case, because it has also a
highest peak located exactly at the interface. \\

\hspace*{0.8cm} To reveal more characteristics of the microscopic
interface modes in SQDs, we have further calculated a few other
physical properties of these interface modes. These include the
average vibrational energy, the total vibrational energy, the
total vibrational amplitudes, and the radial projection of the
vibrational amplitude of the $l$-th shell. These properties are
defined as the following: for the $i$-th mode, the average
vibrational energy of the $l$-th shell,
$E^{i}_{\it l}$, is define as
\begin{math}
E^{i}_{\it l}=m_{\it l}\widetilde{A_l^i}=m_{\it l}\left
[\frac{1}{n}\sum\limits_{k=1}^n
(a_{lk}^i)^{2}\right ]^{\frac{1}{2}}
\end{math}; where $a_{lk}^i$ and $n$ are the same as defined above.
The total vibrational energy of each shell, E$^{i}_{{\it l},T}$,
is defined as
\begin{math}
E^{i}_{{\it l},T}=E^{i}_{\it l}\cdot n
\end{math};
The total vibrational amplitude of each shell in the $i$-th mode is
defined as
\begin{math}
A_{{\it l},T}^{i}=A_{\it l}^{i}\cdot n
\end{math},
where $A_{\it l}^{i}$
is the AVA defined before. The radial projection of the vibrational
amplitude in the ${\it l}-th$ shell is defined by
\begin{math}
A_L^i/A^i=\frac{1}{n}\sum\limits_{k=1}^n |{\bf a_k^i}
\cdot {\bf r_k}|/\left (|r_k|\cdot |a_k^i|\right)
\end{math},
where $\bf a^i_k$ denotes the vibrational amplitude of the $k$-th atom
in the $i$-th mode, and $\bf r_k$ is the position vector of the $k-$th
atom relative to the center of the QD. The quantity $A_L^i/A^i$ describes
to what extent a phonon mode is radial-like.

In Fig. 3, these four physical quantities as well as the AVA defined
earlier are plotted as functions of the shell diameters for an
microscopic interface mode with frequency of 294.20 cm$^{-1}$, that is
an A$_{1}$ mode in 44.76 \AA/48.85 \AA SQDs. The five subfigures from
top to bottom are the plots of the average vibrational energy, the
total vibrational energy, the average vibrational amplitude, the
total vibrational amplitude, and the radical projection of
the vibrational amplitude. In Fig. 3, these five physical quantities of
the $l$-th atomic shell for the same phonon mode are plotted as
functions of the shell diameter. These figures are useful to value how
important the corresponding mode is in experiments. It is seen that
the first four quantities all have peaks at the interface of GaAs/AlAs
and the peaks decay with oscillatory manner as the distance from
the interface increases. The fifth subfigure show that for this
A$_1$ microscopic interface modes, the radial projections of the inner
shells are almost equal to 1. In this 44.76\AA/48.85\AA SQD, this is
true for shells with diameter less than 30\AA. The radial
projections of shells close to the interface and surface are fluctuating,
indicating that the vibrational motion of atoms in these shells are
disturbed by the influence of the interface and surface. The average
radial projection of the entire SQD for this A$_1$ mode, $< A_{L}/A >$,
is equal to 0.7829.

To distinguish the interface modes from other modes in SQDs, we
have introduced another useful physical quantity: the k-space
projection strength of SQDs. Suppose that the eigen states of a
phonon mode in the SQDs can be written as
$\phi_{\alpha,j}$, where $\alpha$=A$_{1}$, A$_{2}$, E, T$_{1}$,
or T$_{2}$, is the symmetry index, and $j$ is the serial number of
this mode in the $\alpha-th$ representation.
On the other hand, the eigen states of bulk phonon modes can
be written as $\chi_{i,{\bf k}}$ where ${\bf k}$ is the wave
vector in $\bf k$-space, and $i$=1 to 6 corresponding to six bulk
phonon states (three optical modes and three acoustic modes).
Then, the k-space projection strength is defined as
\begin{math}
P_{i,{\bf k}}^{\alpha,j}=\mid \left
<\chi_{i,{\bf k}},\phi_{\alpha,j}\right >\mid^{2}
)^{*}\cdot C_{{\it l},c}^{\alpha,j}exp
\end{math}. In Fig. 4(a), we plot the k-space projection strenth of the
microscopic interface mode with frequency $\omega^{T_{2}}$=299.13
cm$^{-1}$, that is a mode with T$_{2}$ symmetry in a 21.70
\AA/26.95 \AA QD. It is seen that
the projection strength of this mode covers a wide range from
$\Gamma$-X-$\Gamma$-L, and it has significant components for
bulk LO, TO$_2$, and LA modes and a slightly weaker components for
bulk TO$_1$, TA$_1$, and TA$_2$ modes. As a comparison, we have also
plotted the k-space projection strength of an internal mode
of the same QD with frequency $\omega$=286.92 cm$^{-1}$ that
falls inside the bulk GaAs band. We can see that even for this small
SQD, the characteristics of a bulk mode is still shown as a sharp peak
in the k-space. In general our results show that for internal modes of
SQDs with larger scale, the k-space projections strength always show
narrow and sharp peak on one or a few bulk modes. Furthermore, the SQD
modes with A$_{1}$ symmetry usually has dominant projection
in bulk longitudinal mode, while the SQD modes with T$_{1}$ symmetry
usually has dominant projection in bulk transverse mode.\\

\hspace*{0.8cm}
In summary, it is pointed out that by employing a microscopic
valence force field model, many microscopic interface
phonon modes with frequencies in the gap between the bulk AlAs
band and the bulk GaAs band can be identified in SQDs with thin
shells. These interface modes are unable to be described by
means of the continuum dielectric model. The average
vibrational amplitude and the average vibrational energy of each
atomic shell show peaks at the GaAs/AlAs interface, and the peak
value decay in oscillatory manner as the its distance from the
interface increases. The k-space projection of the microscopic
interface modes shows diverse distributions for all six bulk modes,
which are a typical characteristics of an interface mode. \\
{\bf Acknowledgements}\\
This research is supported by the National Science
Foundation (DMR9803005 and INT0001313) and the Research Corporation
(CC4381).

\newpage
\mbox{}\\
{}$^{\dagger}$ On leave from the Department of Physics, Nanjing
University, Nanjing 210093, P. R. China.

\newpage
\vfill \eject

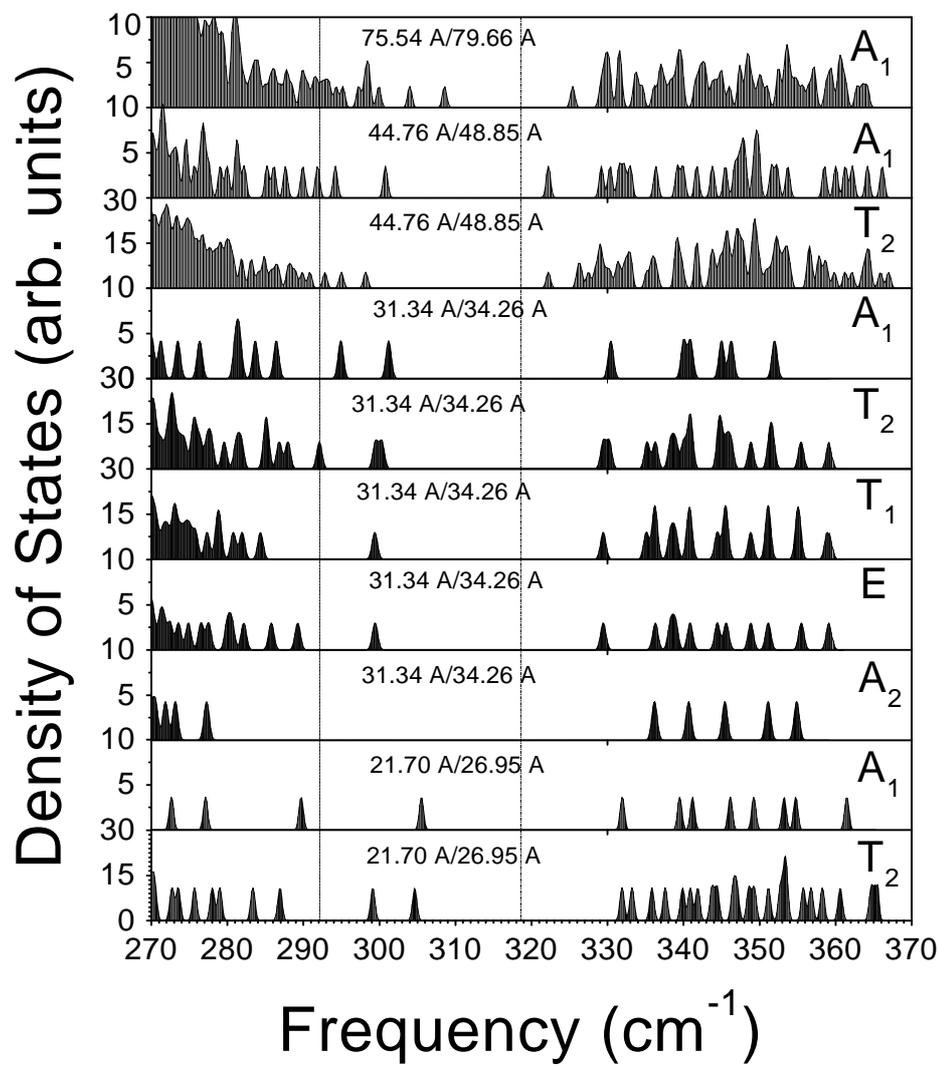
\begin{figure}
\caption{The density of states of phonon modes of selected ten
kinds of SQDs with thin shells}
\label{fig1}
\end{figure}

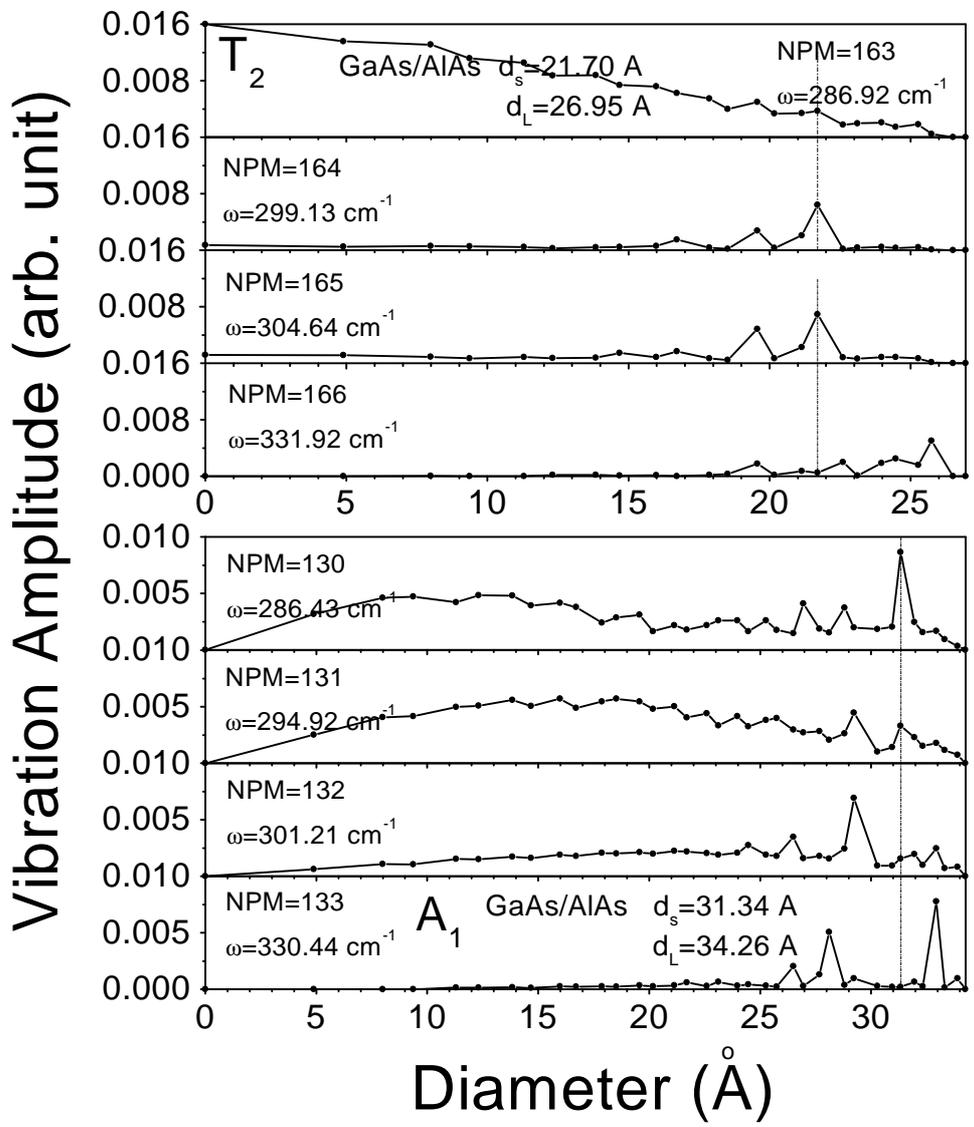
\begin{figure}
\caption{The average vibrational amplitudes as function of
the radius of atomic shells for phonon modes of two kinds of SQDs
with the frequencies covering the whole range from the upper
region of bulk GaAs band to the lower region of bulk AlAs band}
\label{fig2}
\end{figure}

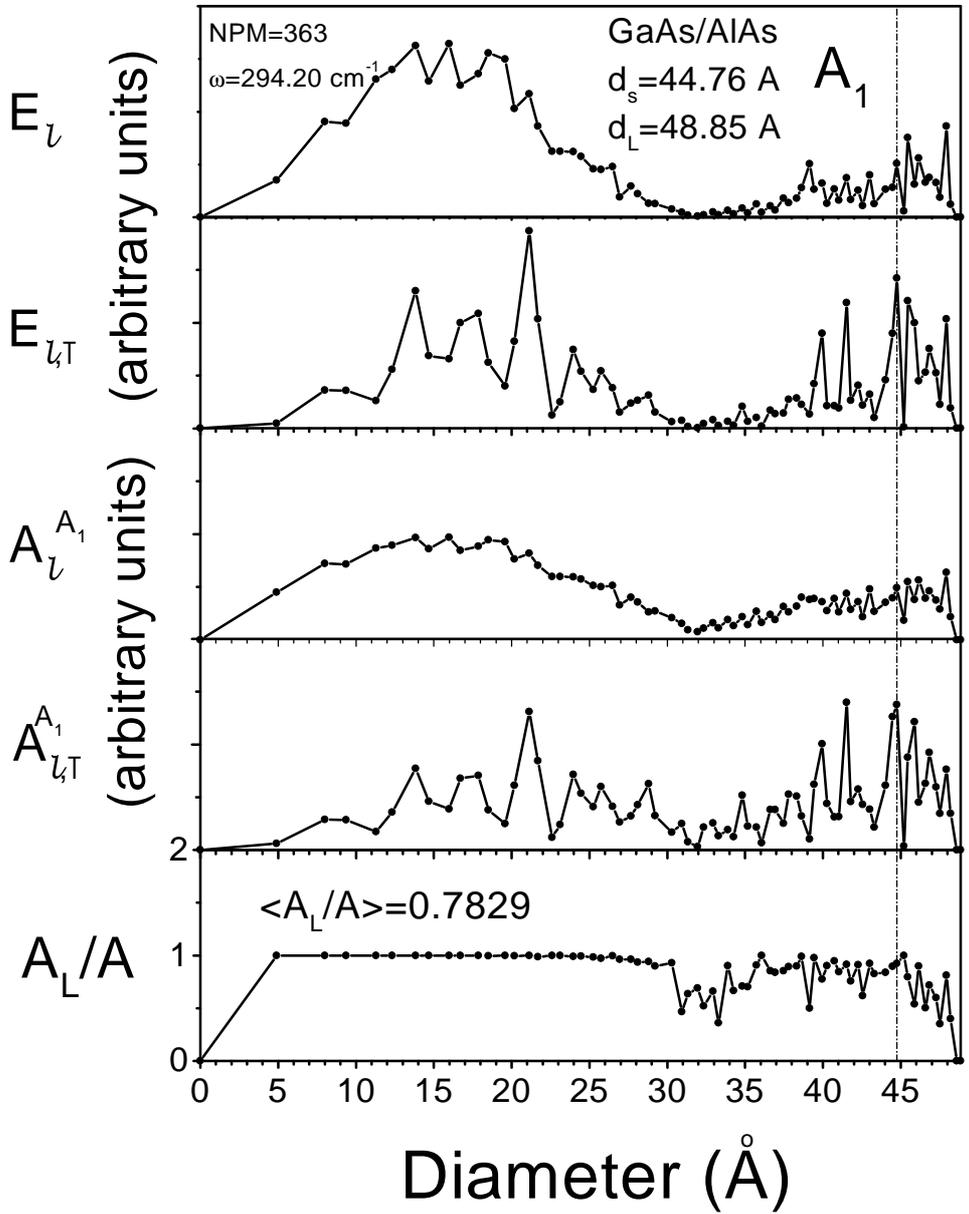
\begin{figure}
\caption{The vibrational energy, the total vibrational energy, the average
vibrational amplitude, the total vibrational amplitude, and
the radial projection of the vibration amplitude as function
of the radius of atomic shells for microscopic interface mode
with frequency 294.20 cm$^{-1}$ of A$_{1}$ symmetry in SQDs with
d$_{s}$=44.76 \AA and d$_{L}$=48.85 \AA}
\label{fig3}
\end{figure}

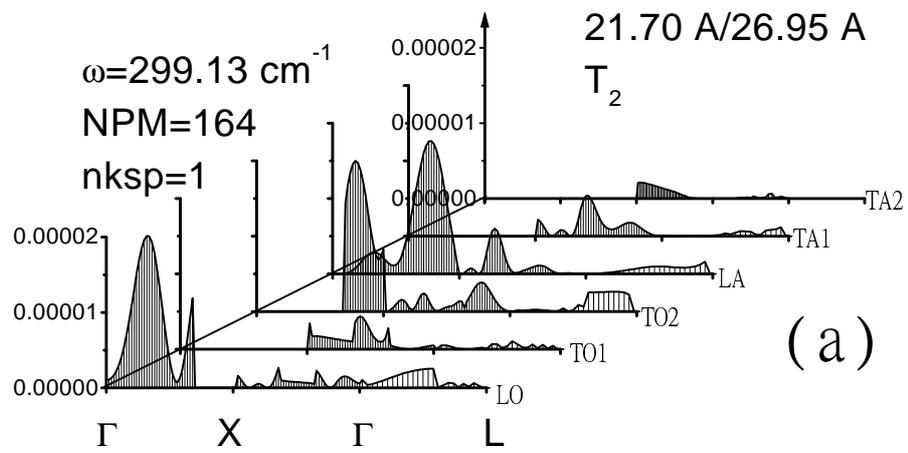
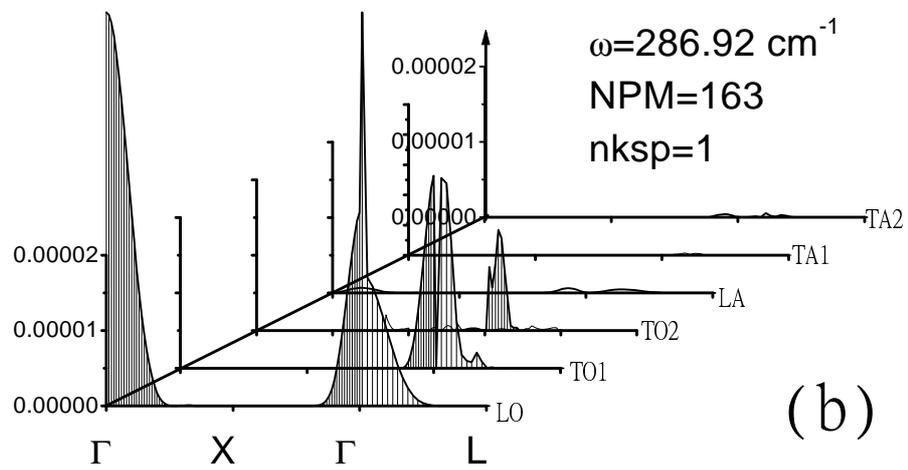
\begin{figure}
\caption{The k-space projection of vibrational amplitudes for
modes of T$_{2}$ symmetry in SQDs with d$_{s}$=21.70 \AA and
d$_{L}$=26.95 \AA.}
\label{fig4}
\end{figure}


\begin{thebibliography}{99}
\bibitem{Armand} G. Armand, L. Dobrzysnski, and P. Masri, J. Vacuum Sci.
Technol. 9, 705 (1971).
\bibitem{Maradudin} R. E. Camley, B. Djafari-Rouhani, L. Dobrzynski,
and A. A. Maradudin, Phys. Rev. B27, 7318 (1983).
\bibitem{Giuliani} G. F. Giuliani and J. J. Quinn, Phys. Rev.
Lett. 27, 7318 (1983).
\bibitem{Quinn} G. Qin, G. F. Giuliani, and J. J. Quinn, Phys.
Rev. B28, 6144 (1983).
\bibitem{Cardona} E. Roca, C. Trallero-Giner and M.
Cardona, Phys. Rev. B49,13704 (1994).
\bibitem{Ruppin} R. Ruppin, Phys. Rev. B26, 3440 (1982).
\bibitem{Ruppin2} R. Ruppin, Physica A 178, 195 (1991).
\bibitem{Reinecke} P. A. Knipp and T. L. Reinecke, Phys.
Rev. B46, 10310 (1992).
\bibitem{Lambin} Ph. Lambin, P. Senet, and A. A. Lucas,
Phys. Rev. B44, 6416 (1991).
\bibitem{Ren} S. F. Ren, Z. Q. Gu and D. Y. Lu. Solid
State Communication, 113, 273-277 (2000).
\bibitem{Ren2} S. F. Ren, D. Y. Lu, and G. Qin, Phys. Rev. B63(19),
195315 (2001).
\bibitem{Jap} G. Qin and S. F. Ren, J. of Appl. Phys., (in print, 2001).
\bibitem{Kunc} K. Kunc, M. Naalkanski, and M. Nusimovici,
Phys. Stat. Sol. (b) 72, 229 (1975); K. Kunc, Ann. Phys. (France) 8, 319
(1973-1974).

\bibitem{Ren3} S. F. Ren, H. Y. Chu, and Y. C. Chang, Phys. Rev. Lett.
59(16), 1841 (1987).

\bibitem{Ren4} Optical phonons in GaAs/AlAs quantum wires, S. F. Ren and
Y. C. Chang, Phys. Rev. B  43,  11857 (1991).

\bibitem{Harrison} Harrison, Electronic Structure and the Properties of
Soilds, Freeman, San Francisico, 1980.

\bibitem{Kunc} K. Kunc, M. Naalkanski, and M. Nusimovici, Phys. Stat.
Sol. (b) 72, 229 (1975); K. Kunc, Ann. Phys. (France) 8, 319 (1973-1974).

\bibitem{SYR} S. Y. Ren, Phys. Rev. B55, 4665 (1997).

\bibitem{SYR2} S. Y. Ren, Solid State Comm. 102, 479 (1997).

\bibitem{SYR3} S. Y. Ren, Jpn. J. Appl. Phys, 36, 3941 (1997).

\end{thebibliography}
\end {document}